\documentclass[a4paper,twocolumn,11pt,accepted=2026-04-20]{quantumarticle}
\pdfoutput=1

\usepackage{amsmath,braket,amssymb,graphicx,float,amsthm,xcolor,outlines,mathrsfs,multirow,hhline,bm}
\usepackage[numbers]{natbib}

\DeclareMathOperator{\tr}{Tr}
\usepackage[english]{babel}
\usepackage[utf8]{inputenc}
\usepackage[T1]{fontenc}
\usepackage{listings}
\usepackage[colorinlistoftodos]{todonotes} 
\usepackage[colorlinks=true, allcolors=blue]{hyperref}
\usepackage[caption=false]{subfig}
\newcommand{\nb}{\bar{n}}

\begin{document}
\title{Nonclassical nullifiers for quantum hypergraph states}
\author{Abhijith Ravikumar}
\email{abhijith.ravikumar@upol.cz}
\affiliation{Department of Optics, Palack\'{y} University, 17. listopadu 1192/12, 779 00 Olomouc, Czech Republic}
\author{Darren W. Moore}
\email{darren.moore@upol.cz}
\affiliation{Department of Optics, Palack\'{y} University, 17. listopadu 1192/12, 779 00 Olomouc, Czech Republic}
\author{Radim Filip}
\email{filip@optics.upol.cz}
\affiliation{Department of Optics, Palack\'{y} University, 17. listopadu 1192/12, 779 00 Olomouc, Czech Republic}

\begin{abstract}
Quantum hypergraph states form a generalisation of the graph state formalism that goes beyond the pairwise (dyadic) interactions imposed by remaining inside the Gaussian approximation. Networks of such states are able to achieve universality for continuous variable measurement based quantum computation with only Gaussian measurements. For normalised states, the simplest hypergraph states are formed from $k$-adic interactions among a collection of $k$ harmonic oscillator ground states. However such powerful resources have not yet been observed in experiments and their robustness and scalability have not been tested. Here we develop and analyse necessary criteria for hypergraph nonclassicality based on simultaneous nonlinear squeezing in the nullifiers of hypergraph states. We put forward an essential analysis of their robustness to realistic scenarios involving thermalisation or loss and suggest several basic proof-of-principle options for experiments to observe nonclassicality in hypergraph states. 
\end{abstract}

\maketitle

\section{Introduction}

Quantum hypergraph states were introduced for qubits~\cite{qu_encoding_2013,rossi_quantum_2013} as a generalisation of the graph states most famous for their role in quantum computation as cluster states~\cite{raussendorf_one-way_2001}, and have recently been experimentally realised~\cite{huang_demonstration_2024}. As regards computation, they have the striking property that they reduce the Clifford complexity of measurements required for universality~\cite{miller_hierarchy_2016} which is compensated for in the more challenging preparation of the hypergraph state. A similar property holds for continuous variable (CV) systems, where computational universality can be achieved using only Gaussian measurements for encoded qubits~\cite{baragiola_all-gaussian_2019}. For hypergraph states in a fully CV context this property also holds, in the sense that a network composed of ideal order 3 hypergraph states can implement universal computation for CV using only Gaussian measurements~\cite{moore_quantum_2019}. Such states form a generalisation of the fully Gaussian graph states~\cite{menicucci_graphical_2011,vandre_graphical_2025}. Hypergraph states are directly constructed from higher order multimode nonlinearities, replacing the pairwise (dyadic) Gaussian interactions used in the formal construction of cluster states~\cite{menicucci_universal_2006} with triadic and higher order interactions. As a consequence the nonlinear features of the states are deeply embedded in their intermodal features and multipartite quantum correlations. 

The nonclassicality of these correlations for physical, finite energy, hypergraph states can be captured by simultaneous nonlinear squeezing~\cite{moore_hierarchy_2022} in the hypergraph state nullifiers. This is a dual generalisation of both the nonlinear squeezing derived from single-mode nonlinear phase gates and used for nonlinear feedforward~\cite{sakaguchi_nonlinear_2023}, and the simultaneous linear squeezing of nullifiers used to verify standard cluster states~\cite{yokoyama_ultra-large-scale_2013,chen_experimental_2014,yoshikawa_invited_2016,asavanant_generation_2019,larsen_deterministic_2019}. Recently the use of nullifiers has also been extended to the verification of non-Gaussian cluster states~\cite{walshe_streamlined_2021,banic_exact_2025,kala_nullifiers_2025,ostergaard_octo-rail_2025,hosseinynejad_realistic_2026}. The simultaneous nonlinear squeezing must occur among nonlinear combinations of quadrature operators from different modes. As these correlations are spread across many modes, the depth with respect to various noise processes is a nontrivial property. The sensitivity of nonclassicality evaluation to initial thermal noise is intuitive: increasing thermal noise destroys the nonclassicality, but can be compensated for by initial Gaussian squeezing or increased nonlinear coupling strength. However sensitivity to both loss and thermalisation bear a complex relationship to the nonlinear coupling strength and the Gaussian squeezing of the initial state. Both the nonlinear coupling strength and the initial squeezing must be optimised with respect to the loss/thermalisation to maximise the nonclassical depth. Surprisingly, in some cases momentum squeezing can increase sensitivity to loss/thermalisation, contrary to the ideal case. These essential robustness analyses motivate us to provide examples of current experimental platforms most likely capable of observing hypergraph nonclassicality.

\begin{figure*}[t]
    \centering
    \includegraphics[width=0.65\columnwidth]{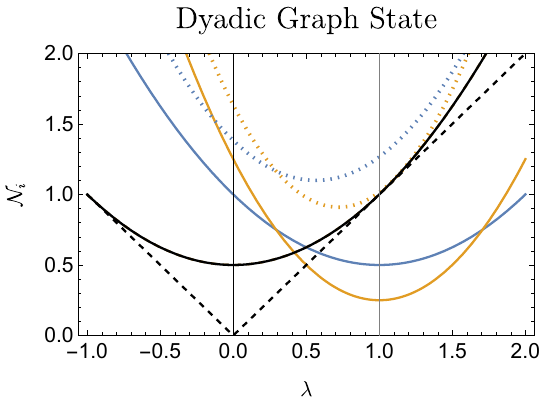}
    \includegraphics[width=0.65\columnwidth]{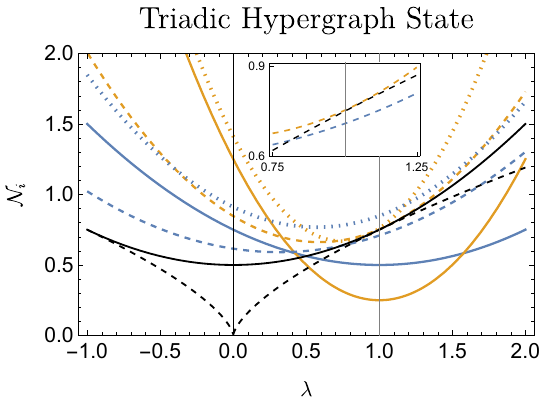}
    \includegraphics[width=0.65\columnwidth]{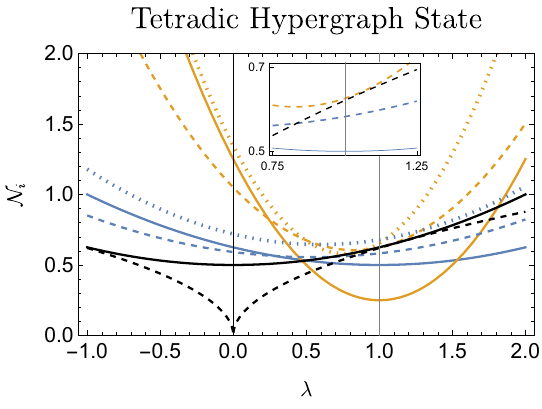}
    \caption{The nonclassical nullifier variance for the dyadic (Gaussian cluster state), triadic and tetradic hypergraph states with $\gamma=1$ (vertical line). They are prepared with no initial squeezing ($r=0$) (blue) or 3~dB momentum squeezing (yellow). The values for initial loss (dashed, $T_{\text{Dyad}}=0$, $T_{\text{Triad}}=0.46$, and $T_{\text{Tetrad}}=0.73$), or thermalisation (dotted, $\bar{n}_{\text{Dyad}}=0.38$, $\bar{n}_{\text{Triad}}=0.16$ and $\bar{n}_{\text{Tetrad}}=0.09$) are the nonclassical depths for the unweighted hypergraph state with initial 3~dB momentum squeezing. Nonclassicality is evidenced by the squeezing of $\mathcal{N}_i$ below the level of the ground state (black). Furthermore the nullifier nonclassicality is not due to the initial Gaussian squeezing of the states, as the nonlinear squeezing goes below the variance achievable with local squeezing (dashed black). For pure states (solid), initial squeezing in momentum always increases the nonlinear squeezing of the nullifier at $\lambda=\gamma=1$. For the hypergraph states loss and thermalisation counteract initial squeezing in momentum such that the initial squeezing must be optimised with respect to the hypergraph weight $\gamma$ and the type and strength of the loss or thermalisation. This can be seen in the insets at $\lambda=1$ where the fixed initial squeezing results in greater sensitivity to loss (yellow above blue).
    }
    \label{Nullifier}
\end{figure*}

\section{Hypergraph Nonclassicality}

The simplest $k$-uniform hypergraph states~\cite{moore_quantum_2019} are those of unit weight, composed of $k$ modes to which are applied a $k$-adic interaction. We refer to $k$-mode and $k$-uniform hypergraph states as triads, tetrads, etc. These are the (unnormalised) states
\begin{equation}\label{IdealHG}
    \ket{\Psi}=e^{i\prod_j^kq_j}\ket{p=0}^{\otimes k}\,
\end{equation}
where $[q_j,p_j]=i$ are the canonical position and momentum, and $\ket{p=0}$ is the zero-momentum eigenstate of the momentum operator. Such eigenstates have infinite energy, therefore Eq.~(\ref{IdealHG}) can only be approached asymptotically. The hypergraph operator $e^{i\prod_jq_j}$ can be seen as a generalisation of the two-mode Gaussian CZ gate used to construct cluster states. The ideal unphysical limit of these states can be represented by the nullifier formalism~\cite{gu_quantum_2009,moore_quantum_2019}, in which $\ket{\Psi}$ is uniquely identified by the collection of commuting operators $\{N_i=p_i-\prod_{j\ne i}q_j\}$. More precisely, $\ket{\Psi}$ is stabilised by the operator $e^{iN_i}$, with the meaning that $e^{iN_i}\ket{\Psi}=\ket{\Psi}$. It follows that $N_i\ket{\Psi}=0$. More physically, we may construct the state
\begin{equation}
    \ket{\psi}=e^{i\gamma\prod_j^{k} q_j}\ket{r}^{\otimes k}\,,
    \label{SqueezedkHypergraphstateeq}
\end{equation}
where $\gamma$ is the strength of the nonlinear interaction, $\ket{r}$ is the squeezed state $\ket{r}=S(r)\ket{0}$, with $S(r)=e^{ir(qp+pq)}$ and $\ket{0}$ the harmonic oscillator ground state. For clarity, we will refer to $\ket{\Psi}$ as the ideal hypergraph state and $\ket{\psi}$ as a physical hypergraph state. For a weighted ideal hypergraph state, the nullifiers are modified to the set $\{N_i=p_i-\gamma \prod_{j\ne i}q_j\}$, where we observe that the weight $\gamma$ is the same for each nullifier and corresponds to the nonlinear strength of the hypergraph operator. 

For graph states produced in experiments, the variance of the nullifiers forms a test for the kind of correlations embodied by the ideal states~\cite{yokoyama_ultra-large-scale_2013,chen_experimental_2014,yoshikawa_invited_2016,asavanant_generation_2019,larsen_deterministic_2019}. Here we follow the same pattern and define the set of nullifier variances
\begin{equation}
    \mathcal{N}_i=\text{Var}\left(p_i+\lambda\prod_{j\ne i}q_j\right)\,,\label{NullV}
\end{equation}
where $\lambda$ is a hypergraphicity parameter which identifies the weight of the hypergraph. In the case of minimal information about the quantum state, even only the information on the nullifier variances, the hypergraphicity is the value of $\lambda$ which operationally identifies the effective hypergraph weight $\gamma_\text{eff}$ by attaching the ideal model in Eq.~(\ref{SqueezedkHypergraphstateeq}) to the nullifier variances. In the realistic case of unequal $\mathcal{N}_i$ the value of $\gamma_\text{eff}$ can be chosen by averaging over the $\mathcal{N}_i$, due to the symmetry around the $k$ nullifiers. Alternatively a given hypergraph weight may be targeted ($\lambda=\gamma$) and then the hypergraphicity identifies whether or not nonclassicality in the target hypergraph state has been accomplished. 

Noise reduction in such nonlinear combinations of variables is a nonlinear squeezing which carries information about the non-Gaussian properties of the state~\cite{moore_hierarchy_2022}. Therefore these nullifier variances are a dual generalisation of the graph state verification method to hypergraph states, and of nonlinear squeezing to multimode nonlinear squeezing. Simultaneous squeezing in the collection of $\mathcal{N}_i$ is a synthesis of both ideas targeting the statistical and directly measurable properties induced by a multimode nonlinearity. Importantly, for physical hypergraph states we still require that the nullifiers are uncorrelated, $\text{Cov}(N_i,N_ j)=0$, and that the mean values are zero, $\braket{N_i}=0$.

Each $\mathcal{N}_i$ alone is able to form a basic criterion for nonclassicality~\cite{moore_estimation_2019}. In order to see this, first observe that evaluating these variances on the $k$-mode coherent state $\ket{\bm{\alpha}}=\otimes_i\ket{\alpha_i}$, with $\bm{\alpha}\in\mathbb{C}^k$, produces
\begin{equation}\label{CohNullVar}
    \mathcal{N}_i=\frac12+\lambda^2\left(\prod_{j\ne i}\left(\frac12+2\Re(\alpha_j)^2\right)-\prod_{j\ne i}\left(2\Re(\alpha_j)^2\right)\right)\,,
\end{equation}
where $\Re$ 
denotes the real 
part of the $\alpha_j$. This expression is bounded from below by the ground state. In order to form a threshold for nonclassicality, the ground state variance should also be a lower bound for arbitrary incoherent mixtures of coherent states. We sketch the proof here and leave the detailed argument to Appendix~\ref{A}. These variances are independent of displacements in momentum i.e. independent of $\Im(\bm{\alpha})$. Therefore any state can be displaced in momentum without altering the value of $\mathcal{N}_i$. It follows that each state can be replaced by a state with the same value of $\mathcal{N}_i$ but where the first moment is zero, so that $\mathcal{N}_i=\braket{\left(p_i+\lambda\prod_{j\ne i}q_j\right)^2}$, which is a linear function of the state. The pure states are the extremal states so incoherent mixtures of the coherent states are bounded by the pure coherent states, which are bounded by the ground state. This means precisely that a state with noise in the nullifier below the level of the ground state cannot be represented as any mixture of coherent states, which is the definition of $P$-function nonclassicality. 

\begin{figure*}[t]
    \centering
    \includegraphics[height=4cm,keepaspectratio]{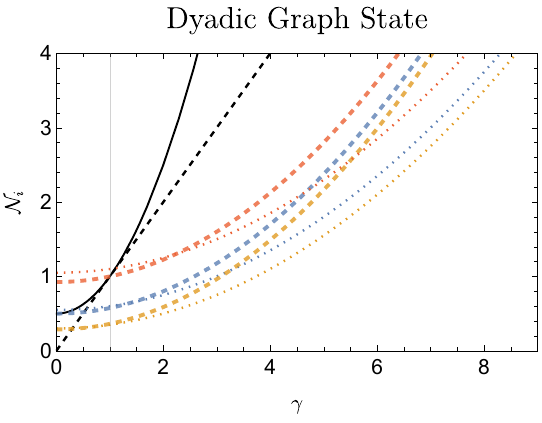}
    \includegraphics[height=4cm,keepaspectratio]{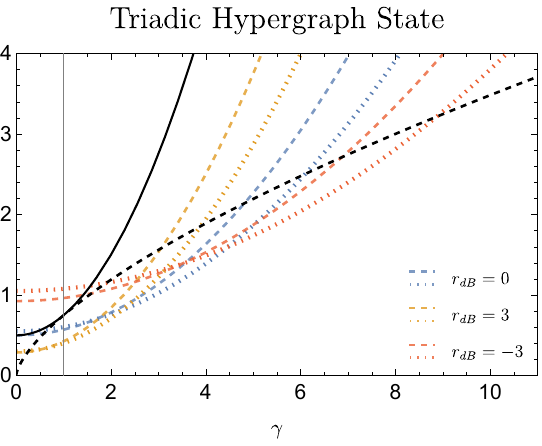}
    \includegraphics[height=4cm,keepaspectratio]{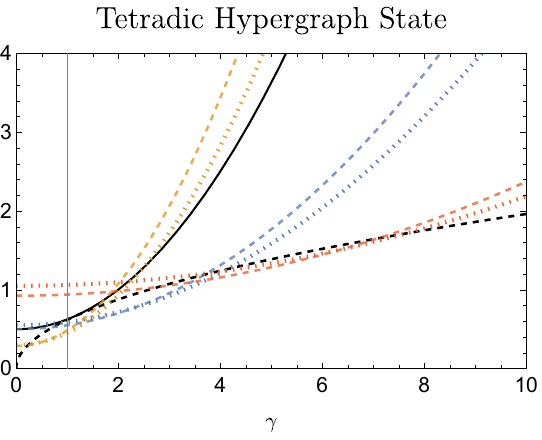}
    \caption{The minimal required interaction strength $\gamma$ to reach nonclassical nullifier variance, aided by initial squeezing and in the presence of small loss or thermalisation. The hypergraphicity is chosen to target the pure state nullifiers at $\lambda=\gamma$. The values for loss $T=0.85$ (dashed curves) and thermalisation $\nb=0.05$ (dotted curves) and for position squeezing (red), momentum squeezing (yellow), or no squeezing (blue), are selected to be illustrative (but see also Appendix~\ref{D}). Nullifier variances below the ground state variance (black) and locally squeezed states threshold (dashed black) indicate nonclassicality. For the Gaussian dyad momentum squeezing is always beneficial, and position squeezing always harmful, regardless of loss or thermalisation. In distinct contrast, the response to loss and thermalisation for the non-Gaussian triad and tetrad depends on $\gamma$. As $\gamma$ increases the nonclassical depth of the nullifiers depends on the type of initial linear squeezing. For low $\gamma$, including the unweighted hypergraph state $\gamma=1$, momentum squeezing increases tolerance to loss and thermalisation compared to position squeezing. For larger $\gamma$ this behaviour transitions to a negative effect, where nonclassicality is lost faster than with no squeezing. Instead in this regime position squeezing increases the tolerance to loss and thermalisation, despite having lower nonlinear squeezing at the level of the pure states.}
    \label{LossATherm}
\end{figure*}

If any $\mathcal{N}_i$ is squeezed below the level of the ground state for any $\lambda$ then the state is nonclassical. This is as yet insufficient to begin connecting the state with a particular weighted $k$-uniform hypergraph state. All $k$ nullifiers may be made nonclassical merely through local squeezing on each mode. A stronger condition can be composed which excludes the possibility that such nonclassical statistical properties could be achieved by a collection of uncorrelated Gaussian squeezed states (see Appendix~\ref{B}). All $k$ nullifiers are required to simultaneously show nonlinear squeezing below that available to uncorrelated squeezed states. In order to connect to hypergraphs, we further require identical $\lambda$ for each nullifier variance. We refer to such a state of affairs as hypergraph nonclassicality. Then the hypergraphicity $\lambda$ identifies which effective hypergraph state with weight $\gamma_\text{eff}$ is targeted most strongly by the nonclassical nullifiers.

The physical hypergraph state $\ket{\psi}$ bears the simplifying symmetry that the $k$ nullifier variances are equal. Therefore for the analysis of this state the nonclassicality for one nullifier variance implies hypergraph nonclassicality. In general, when the state is not known, all $k$ nullifier variances must be evaluated independently. For $\ket{\psi}$ we have
\begin{equation}\label{HGVar}
    \mathcal{N}_i=\frac{e^{2r}}{2}+\frac{(\lambda-\gamma)^2 e^{-2r(k-1)}}{2^{k-1}}\,,
\end{equation}
which shows nonclassicality for any $\gamma\ne0$, demonstrated for the nonsqueezed triad and tetrad (blue and yellow solid curves) in Fig.~\ref{Nullifier}. In fact this expression never goes below $\frac{e^{2r}}{2}$ for any $\lambda$ or $\gamma$. Instead, $\gamma\ne0$ results in a lateral translation of the $\lambda$ parabola in the plane so that the minimal nonlinear squeezing is at $\lambda=\gamma$. The $k$-adic interaction generates the correlation $\text{Cov}\left(p_i,\prod_{j\ne i}q_j\right)$, as can be seen by expanding Eq.~(\ref{NullV}) in moments~\cite{moore_estimation_2019}.

For pure states and unitary processes nonlinearities which are functions of only $q_i$ or $p_i$ can be arbitrarily strengthened by prior Gaussian squeezing (Fig.~\ref{Nullifier}, yellow solid curves), resulting in unlimited nonlinear squeezing. The challenge is that the response of these nullifier variances to realistic loss and noise processes depends strongly and counterintuitively on the particular process, the finite squeezing $r$, the nonlinear strength $\gamma$, and on the number of modes. The point is illustrated in Fig.~\ref{Nullifier} by the different effects on the nullifier nonclassicality by loss or thermalisation (dashed and dotted curves respectively). As expected, for the Gaussian dyadic graph state (equivalent to the smallest cluster state), nonclassicality is maintained up to full linear loss ($T=0$). This is not the case for the higher polyads. While initial momentum squeezing for pure states increases the nonlinear squeezing, it also increases sensitivity to the initial impurity as can be seen from the loss curves (blue and yellow dashed) for the triad and tetrad. Therefore, the initial squeezing $r$ must be optimised against $\gamma$ and $T$ or $\nb$ in order to maximise the nullifier nonclassicality over $\lambda$. Then the hypergraphicity $\lambda$ corresponding to the largest robustness to the main imperfection (loss or noise) operationally specifies the weight of the hypergraph state alongside the value of the nullifier variance with respect to the classical threshold. Examples can be seen in the centre and right panels of Fig.~\ref{Nullifier}, where loss or thermalisation alters the minimal nonlinear squeezing away from $\gamma=1$.

Motivated by two different experimental platforms suitable for small hypergraph states, superconducting circuits and trapped ions, we analyze the dominant error mechanisms: circuit radiation loss~\cite{li_autonomous_2024} and ion mechanical thermalization~\cite{podhora_quantum_2022}. The loss mechanism involves exposing each mode of the hypergraph state to a purely lossy channel of energy transmittivity $T$ with the oscillator ground state in the second port. The thermalisation mechanism subjects each mode of the quantum state to random displacements in phase space, characterised by a thermal occupation $\nb$. The thermalisation map on a state $\rho$ is $M_{\nb}(\rho)=\int\frac{d^2\alpha}{\pi\nb}e^{-\frac{|\alpha|^2}{\nb}}D(\alpha)\rho D^\dagger(\alpha)$~\cite{podhora_quantum_2022}. The nullifier variances of Eq.~(\ref{HGVar}) are thus modified to
\begin{multline}\label{HGVarL}
  \mathcal{N}_i=  \frac{1+T(e^{2r} -1)}{2} +\lambda^2 \left(\frac{1+T(e^{-2r} -1)}{2}\right)^{k-1} \\+\frac{e^{-2r(k-1)}}{2^{k-1}}\gamma\left(T \gamma -2 \lambda  T^{\frac{k}{2}}\right)\,,
\end{multline}
for loss and 
\begin{multline}\label{HGVarTh}
  \mathcal{N}_i= \frac{e^{2 r}}{2} +\bar{n} +\lambda ^2 \left(\bar{n}+\frac{e^{-2 r}}{2}\right)^{k-1}\\ +\frac{e^{-2r(k-1)}}{2^{k-1}} \gamma(\gamma-2\lambda)\,,
\end{multline}
for thermalisation. It can be seen directly from Eqs.~(\ref{HGVarL}) and (\ref{HGVarTh}) that loss $T$ and thermalisation $\nb$ both interact differently and nontrivially with the initial squeezing $r$ and nonlinear strength $\gamma$.

To analyse this counterintuitive behaviour for the physical states $\ket{\psi}$, we set $\lambda=\gamma$ which is the condition for maximal nonlinear squeezing for the pure hypergraph states. There is a careful point to distinguish here. After exposure to noise the hypergraphicity $\lambda=\gamma$ typically no longer corresponds to the absolute minimum of the parabola and therefore no longer identifies the effective hypergraph weight of $\ket{\psi}$ . This choice instead fixes the nonlinear variances to the nullifier corresponding to the pure hypergraph state i.e. we analyse how the nullifier variances of these states are degraded by noise. Operationally, in the absence of the knowledge of the state exposed to noise, the new minimum of the parabola specifies a new hypergraphicity. Bear in mind that nullifiers only specify the exact state when the operator is equal to zero. Nullifier variances have a much smaller resolution than this strict requirement, particularly for non-Gaussian states where first and second moments are insufficient to characterise the quantum state.

\begin{figure}[h]
    \centering
    \includegraphics[width=\columnwidth]{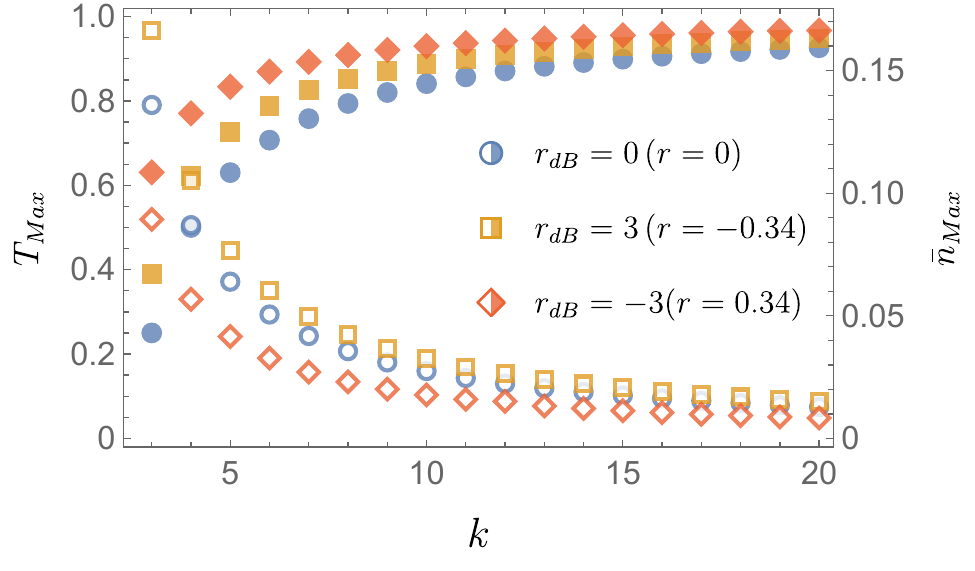}
    \caption{Nonclassicality depths of hypergraph states. Greatest amount of loss $T_{Max}$ (full plot markers) and  greatest thermal occupation number $\bar{n}_{Max}$ (empty plot markers) still showing hypergraph nonclassicality against hypergraph order (number of modes $k\ge3$) for momentum (yellow), position (red) squeezing or no (blue) squeezing, and optimal nonlinear strength $\gamma$. The extra robustness granted due to initial squeezing decreases as the number of modes increases.}
    \label{kModesLoss}
\end{figure}

\section{Hypergraph Nonclassicality Under Noise Processes}

As shown qualitatively in Fig.~\ref{Nullifier}, the nonclassicality of the nullifiers for the triad and tetrad does not respond to loss or thermalisation as for the dyad. Increasing initial momentum squeezing actually compromises the robustness of the nullifier variances $\mathcal{N}_i$ to the noise processes (see insets in Fig.~\ref{Nullifier}). Due to this, the initial squeezing must be optimised with respect to the nonlinear strength $\gamma$ and the loss or thermalisation. The optimisation can efficiently compensate for some negative effects of loss or thermalisation, as is the case with the fully Gaussian dyad. Importantly the response of the hypergraph states to initial thermal noise is not a limiting factor as it can always be compensated for by initial Gaussian squeezing (see Appendix~\ref{C}).

Fig.~\ref{LossATherm} shows the response of the nonlinear squeezing to noise as a function of $\gamma$ for small loss or thermalisation. The dyad case illustrates the standard behaviour of linear systems: the squeezing of the nullifier variance of the two-mode state generated by $e^{i\gamma q_1q_2}$ at $\lambda=\gamma$ under noisy conditions is ordered, from least to most, by position squeezing (red), no squeezing (blue) and momentum squeezing (yellow). This holds across all values of $\gamma$. However for the non-Gaussian triad and tetrad this ordering is interrupted beyond small values of $\gamma$, near $\gamma=2$ and $\gamma=1.8$ respectively. Here the interaction of the noise with the initial momentum squeezing increases the sensitivity to the noise such that greater robustness is achieved with zero squeezing. As $\gamma$ is increased further this ordering further shifts so that initial position squeezing results in greater robustness. This occurs in spite of the lower nonlinear squeezing achieved when there is no exposure to noise. 

This holds not just for these fixed values of $T$ and $\nb$, but in general for the maximum sustainable values before hypergraph nonclassicality is lost (see Appendix~\ref{D}). For the Gaussian dyad momentum squeezing is always the best strategy to increase tolerance to loss or thermalisation. Furthermore this strategy can always compensate for loss or thermalisation by making the two-mode squeezing visible again, even if the quantity of squeezing is reduced. In contrast, for the non-Gaussian hypergraph states a given value of $\gamma$ requires an optimisation of the initial squeezing to maximise the tolerance of hypergraph nonclassicality to loss and thermalisation. Moreover this optimal strategy is still insufficient to always compensate for loss or thermalisation by making the hypergraph nonclassicality visible again. That is, for certain values of $T$ or $\nb$ there is no initial squeezing which will recover visibility of the hypergraph nonclassicality. To summarise, preparation of robust non-Gaussian hypergraph states with a fixed hypergraphicity requires an optimisation of the initial squeezing to maximise the tolerance of hypergraph nonclassicality to loss and thermalisation. For the standard unweighted hypergraph states in Fig.~\ref{LossATherm} it is still better to have initial momentum squeezing, although for values of $\gamma>1$ this optimisation transitions through zero squeezing to position squeezing, and for different values of $T$ and $\nb$ the optimal value of $r$ also changes.

As the optimal initial squeezing transitions from momentum squeezing to position squeezing, taking it beyond the unweighted hypergraph states at $\gamma=1$, there is a value of $\gamma$ (see Appendix) where it is beneficial to have zero squeezing. Targeting hypergraphicity close to this value advantageously minimises the extra requirement of initial squeezing from the preparation of hypergraph nonclassicality under noisy conditions. Moreover, across all values of $\gamma$ we find that it is not helpful to use rotated squeezed states (i.e. complex values of $r$). Going from $k=3$ to $k=4$, it can be seen that the response of different kinds of squeezing to loss or thermalisation changes. Fig.~\ref{kModesLoss} shows the robustness of the nonclassical nullifier variances as a function of the order of the hypergraph state $\ket{\psi}$, with an optimised choice of nonlinear strength $\gamma$. Here we are able to see a clear qualitative difference between the effect of loss and thermalisation. For loss, optimising over $\gamma$ shows that it is always better to have zero squeezing, as the solid blue points are always below the red and yellow points. For thermalisation, it is always better to have momentum squeezing as the yellow points are always above the red and blue. This suggests that for higher order hypergraph states platforms that are dominantly affected by loss will be more efficient for state preparation, provided the required nonlinear interactions can be accommodated.

At the start of this discussion we fixed the hypergraphicity $\lambda=\gamma$ to test the effect of loss and thermalisation on the physical states $\ket{\psi}$ with a hypergraph weight $\gamma$. Instead, one may fix a target hypergraphicity, (e.g. $\lambda=1$) and test how changing the conditions of nonlinear strength and loss/thermalisation affect the achievement of this target, while maintaining nullifier variances below the nonclassicality thresholds. For $\lambda=1$ it is still better to have initial momentum squeezing for the small noise values given in Fig.~\ref{LossATherm}. However this changes for both different $\lambda$ and different values of noise, and must similarly be appropriately optimised.

\section{Feasible sources of triadic interactions}

The interactions required in order to implement the hypergraph operators are highly nonlinear and highly multimode, a combination that has not been well explored experimentally for continuous variable systems. Here we detail some plausible platforms which may be used for the first time to directly implement the poly-adic interactions required for hypergraph states. Firstly, with trapped ions trilinear interactions in the rotating wave approximation have been realised in both two ion~\cite{ding_quantum_2017} and three ion~\cite{ding_quantum_2018,maslennikov_quantum_2019} settings. In superconducting circuits such interactions have achieved a third order squeezer~\cite{chang_observation_2020}. In this vein we may approximate the hypergraph operator as $e^{i\gamma q_1q_2q_3}\approx\mathbf{I}+i\gamma q_1q_2q_3$, which when acting on the ground state corresponds to the simultaneous creation of three excitations, a basic effect in the triadic process.
However this state has limited applicability as a hypergraph state, showing nonclassical nullifiers only for $|\gamma|<2.88$. Very recently such systems were used to produce genuine tripartite non-Gaussian entanglement~\cite{jarvis-frain_observation_2025}.

Beyond the rotating wave approximation, triadic interactions arise naturally from the Coulomb interaction between 2 harmonically trapped particles, with equal charge and equal mass, confined along the $z$-axis and with inter-particle distance at equilibrium $z_0$. The third order interaction term in an expansion around the relative positions takes the form $H_3=-\frac{3\kappa}{z_0^4}\Delta z\left(3\left(\Delta x^2+\Delta y^2\right)-2\Delta z^2\right)$ where $\Delta x=x_1-x_2$, $\Delta y=y_1-y_2$ and $\Delta z=z_1-z_2$ are the relative positions of the ions. The induced interaction between orthogonal modes results in terms like $(x_1-x_2)^2\left(z_1-z_2\right)$, whose expansion includes the triadic terms $x_1x_2z_1$ and $x_1x_2z_2$~\cite{marquet_phononphonon_2003,lemmer_quantum_2018}. While most accessible for trapped ion systems, such interactions may also become available for levitated particles, at vastly different mass and frequency scales to ions~\cite{penny_sympathetic_2023}. Ion traps with specially designed geometries can also implement nonlinear intermodal interactions, as used in quantum thermodynamics experiments~\cite{rosnagel_single-atom_2016}. Moreover, for superconducting circuit elements a three-wave mixer known as a SNAIL was proposed~\cite{zorin_josephson_2016,frattini_3-wave_2017} and then implemented~\cite{frattini_optimizing_2018,ezenkova_broadband_2022,eriksson_universal_2024}, and has the capacity to display the required triadic interactions. Proposals for tetradic interactions are, to our knowledge, still theoretical as in the context of superconducting circuits~\cite{kawakami_four-body_2025}. As discussed above, these systems have the advantage that their predominant source of decoherence is loss rather than thermalisation, which our results suggest will minimalise the required initial linear squeezing. 

Beyond these hardware-direct and therefore powerful methods, one may consider that although the standard theoretical construction of cluster states requires the CZ gate, in practice a variety of other methods are used; the entanglement is typically generated by the input of nonclassical (squeezed) states to a beamsplitter. Similar methods may be applied to hypergraph states. The triadic interaction quantum gate can be decomposed into a universal gate set involving Gaussian operations and the cubic phase gate~\cite{kalajdzievski_exact_2021}. Building on this, higher order hypergraphs can also be constructed out of lower ones~\cite{vandre_graphical_2025} or with measurement based methods~\cite{hanamura_implementing_2024}. We leave an analysis of the robustness to noise of such techniques to further work.

\section{Discussion}

Detailed descriptions and realistic analyses of multimode nonlinearities for quantum continuous variables are not well explored experimentally. To remedy this situation we propose a basic nonclassicality criterion linked to the correlations associated with nullifiers of the continuous variable hypergraph states, which we dub hypergraph nonclassicality, inspired by the approach for the Gaussian cluster states. To the best of our knowledge, continuous variable hypergraph states have yet to be produced experimentally. Therefore we provide an essential analysis of the critical robustness of hypergraph nonclassicality against the most common loss and noise processes and show that enhancing the depth of the nonclassical nullifiers in the presence of loss or thermalisation is possible via optimisation. That is, the initial Gaussian squeezing must be chosen in tandem with the nonlinearity $\gamma$ such that the robustness of hypergraph nonclassicality to loss or noise is maximised. In particular, the most basic unweighted hypergraph states under small loss or thermalisation typically require momentum squeezing in order to maximise their robustness to the noise processes, whereas higher weighted hypergraph states typically require position squeezing. While momentum squeezing always enhances the nonlinear strength when the interaction is unitary, the enhancement is also counteracted more strongly by loss and noise at higher values of the nonlinear strength. As the number of modes increases the requirement for initial squeezing lowers when the dominant form of noise is loss, suggesting that platforms that mostly experience loss are better candidates for producing hypergraph nonclassicality. If future systems move towards such large scale nonlinear systems techniques beyond optimisation of linear squeezing, such as error correction, will likely have to be developed to accommodate these scaling relations.

It might be natural to assume these complex noise response properties that change with the number of modes is simply due to a larger number of modes being exposed to noise. However we found that similar behaviour occurs for single mode nonlinear phase states (see Appendix~\ref{E}). That is, the optimisation of the initial squeezing required to maximise the robustness of the nonclassical nullifiers is also required for the nonlinear phase states. Therefore the nonclassicality of the hypergraph states is not more sensitive to loss or thermalisation than that of the nonlinear phase states with respect to nonlinear squeezing. Importantly then, a first pioneering investigation of quantum hypergraph states in laboratories, and their applications in continuous-variable quantum computing can be pursued as the requirements are not far above the already strict requirements of the highly sought nonlinear phase states.

\begin{acknowledgments}
A.R. and R.F. acknowledge the European Union’s HORIZON Research and Innovation Actions under Grant Agreement no. 101080173 (CLUSTEC) and the Quantera project CLUSSTAR (8C24003) of MEYS Czech Republic.  Project CLUSSTAR has received funding from the European Union’s Horizon 2020 Research and Innovation Programme under Grant Agreement No. 731473 and 101017733 (QuantERA). D.M. and A.R. acknowledge support of 25-17472S of the Czech Science
Foundation and the project CZ.02.01.01/00/22\_008/0004649 of the Czech Ministry of Education, Youth and Sport.
\end{acknowledgments}

\section{Data Availability}

No new data were generated for this article. Code supporting these results can be found at~\cite{ravikumar_2026_19813994}.

\bibliographystyle{quantum}
\bibliography{references.bib}

\newpage
\onecolumngrid

\begin{appendix}

\section{Nonclassicality bound for hypergraph nullifiers}\label{A}

The nullifier variances take the form
\begin{equation}
    \mathcal{N}_i=\text{Var}(p_i+\lambda\prod_{j\ne i}q_j)\,.
    \label{SqueezedkHypergraphNullVar}
\end{equation}
Observe first that the nullifier variances are independent of translations in momentum. This is a general property of such combinations of quadrature variables linear in $p$, and can be seen in the example of the coherent states [Eq.~(\ref{CohNullVar})] This means that any given state can be displaced in momentum without altering the value of $\mathcal{N}_i$. In particular, a representative state can always be selected in which $\braket{N_i}=0$ and the variance is equal to the second moment. Therefore it suffices to show that the ground state variance $\frac12+\frac{\lambda^2}{2^{k-1}}$ is a lower bound for the second moment of the nullifier over mixtures of coherent states. Let $\rho_{\bm{\alpha}}=\int P(\bm{\alpha})\ket{\bm{\alpha}}\bra{\bm{\alpha}}d^2\bm{\alpha}$, with $P(\bm{\alpha})>0$ the Glauber-Sudarshan $P$-function, denote an arbitrary mixture over the coherent states. Then, recalling that $P(\bm{\alpha})$ is normalised to unity, the following inequalities hold
\begin{align}
    \frac12+\frac{\lambda^2}{2^{k-1}}&\le\braket{\bm{\alpha}|N_i^2|\bm{\alpha}}\\
    \Rightarrow\int P(\bm{\alpha})\left(\frac12+\frac{\lambda^2}{2^{k-1}}\right)d^2\bm{\alpha}&\le\int P(\bm{\alpha})\braket{\bm{\alpha}|\mathcal{N}_i|\bm{\alpha}}d^2\bm{\alpha}\\
    \Rightarrow\frac12+\frac{\lambda^2}{2^{k-1}}&\le\tr\left(\rho_{\bm{\alpha}}N_i^2\right)\,.
\end{align}
This is the required bound.

\section{Minimum variance limit for hypergraph nullifiers}\label{B}

The hypergraph nullifier variance Eq. \eqref{SqueezedkHypergraphNullVar} evaluated w.r.t $k$-mode squeezed state ($\ket{\phi}=\ket{r}^{\otimes k}$) reaches its minimum when the squeezing parameter $r$ is optimal. At this value of $r$, the minimum variance is given by,
\begin{equation}
    \mathcal{N}_{Min}=4^{\frac{1-k}{k}} (k-1)^{\frac{1-k}{k}} k  |\lambda | ^{\frac{2}{k}}.
\end{equation}

\section{Initial Thermal Noise} \label{C}
\begin{figure}
    \centering
    \includegraphics[width=0.35\columnwidth]{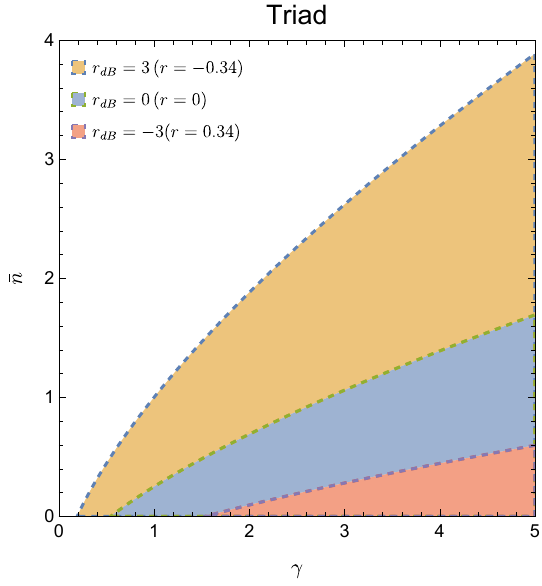}
    \caption{Maximum initial thermal noise $\Bar{n}$ before the hypergraph nonclassicality cannot be distinguished from Gaussian squeezing for the triad. Thermal noise can always be dealt with by increasing momentum squeezing or increasing hypergraph weight $\gamma$.}
    \label{Fig:InitialThermal}
\end{figure}

Initial thermal noise can be introduced into hypergraph states by replacing the squeezed ground state by squeezed thermal states, characterised by a mean occupation $\nb$. The nullifier variance for $k$-mode hypergraph state with this initial thermal noise is given by,
\begin{equation}
    \mathcal{N}_i= \frac{(1+2 \Bar{n})e^{2r}}{2} +\frac{(\gamma-\lambda)^2 e^{-2r(k-1)} (1+2 \Bar{n})^{k-1}}{2^{k-1}}.
\end{equation}
This adds thermal noise directly and simply to each term, otherwise the expression is similar to that of Eq.~(\ref{HGVar}). At the minimum $\lambda=\gamma$ the number of modes has no effect and the thermal noise is easily countered by increased momentum squeezing. Since larger $\gamma$ creates a greater translation of the $\lambda$ parabola with respect to the nonclassicality thresholds, increasing $\gamma$ also counters the negative effect of the thermal noise. These effects are illustrated in Fig.~\ref{Fig:InitialThermal} for the triadic hypergraph state.

\section{Effect of loss and thermalisation on hypergraph nonclassicality} \label{D}

The loss mechanism is modelled by splitting the hypergraph state on a beamsplitter with variable transmittivity $T$, while the loss mechanism is modelled by the thermalisation map defined in the main text and characterised by thermal occupation $\nb$. As detailed, the interaction between initial squeezing $r$, nonlinear strength $\gamma$, and $T$ or $\nb$ is complex and counterintuitive. Here we show that the behaviour we describe in the main text holds at the limits of visible hypergraph nonclassicality beyond local Gaussian squeezed states.

\begin{figure}
        \includegraphics[width=0.24\columnwidth]{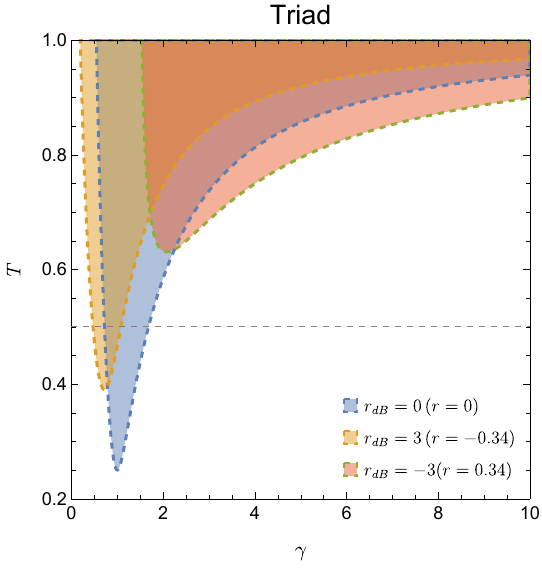}
        \includegraphics[width=0.24\columnwidth]{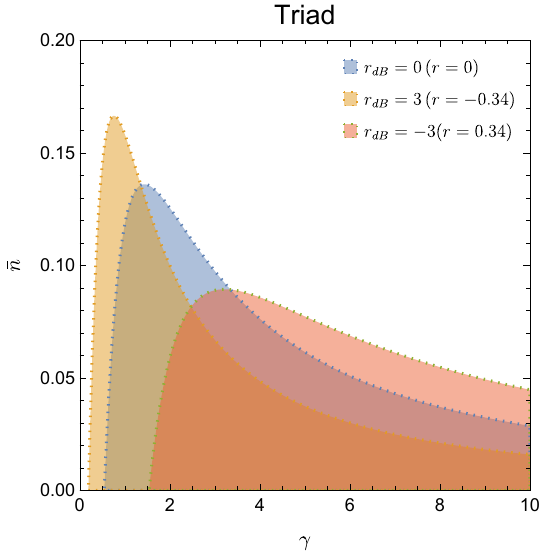}
        \includegraphics[width=0.24\columnwidth]{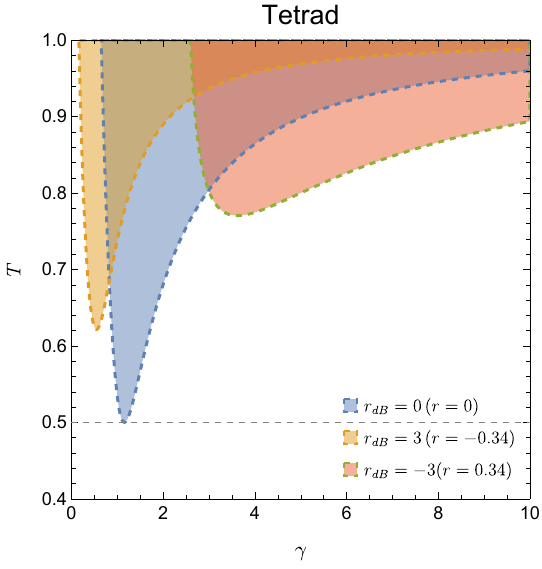}
        \includegraphics[width=0.24\columnwidth]{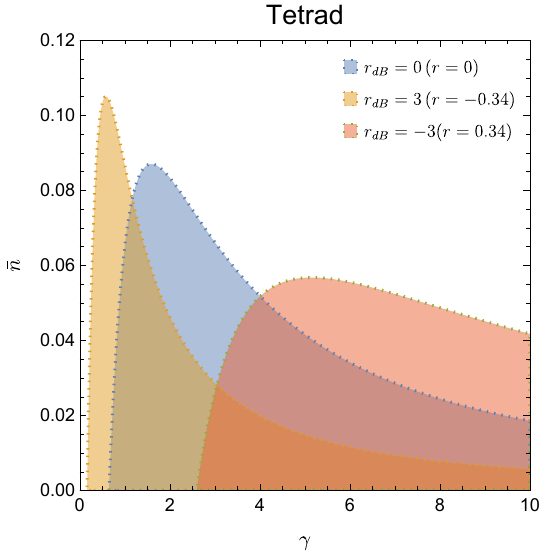}
    \caption{The maximum loss $T$ and thermalisation $\bar{n}$ at the nullifier $\lambda=\gamma$ against the nonlinear strength $\gamma$ for different classes of squeezing. Increasing $\gamma$ beyond an optimal value increases the sensitivity to loss due to attenuation and thermalisation. For values of $\gamma$ lower than this optimal value, loss can be partially compensated for by momentum squeezing. However for larger values of $\gamma$, squeezing also increases the sensitivity to loss. Counterintuitively, for higher values of $\gamma$ position squeezing protects the hypergraph nonclassicality.}\label{FigAppC}
\end{figure}

Fig.~\ref{FigAppC} shows the maximum loss and maximum thermalisation a given hypergraph state can experience while still showing nonlinear squeezing below the threshold of local squeezed states. Increasing the $\gamma$ does not protect against noise, and protection from initial squeezing depends strongly on the $\gamma$. Also visible are the transition from momentum squeezing to position squeezing producing tolerance to the noise processes and the fact that, across all $\gamma$, zero squeezing is better for loss while momentum squeezing is better for thermalisation. 

\section{Comparison with nonlinear phase states} \label{E}

\begin{figure}
    \centering
    \includegraphics[width=0.32\columnwidth]{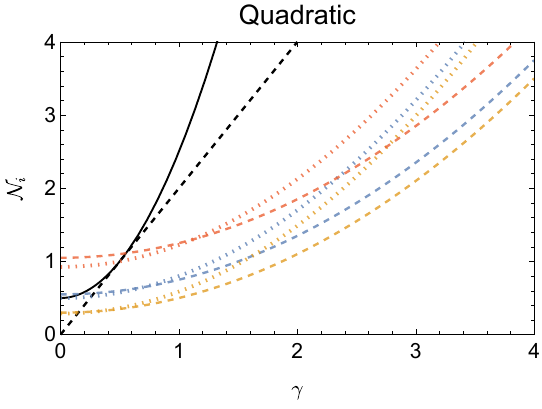}
    \includegraphics[width=0.32\columnwidth]{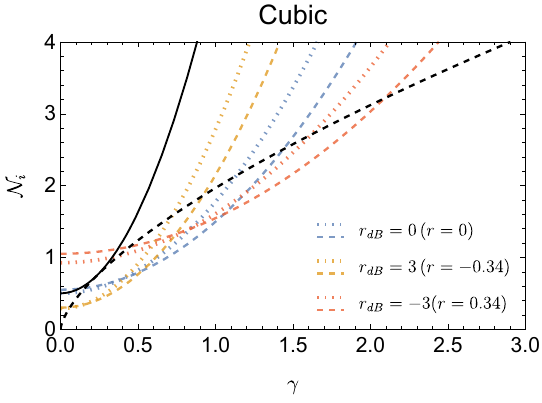}
    \includegraphics[width=0.32\columnwidth]{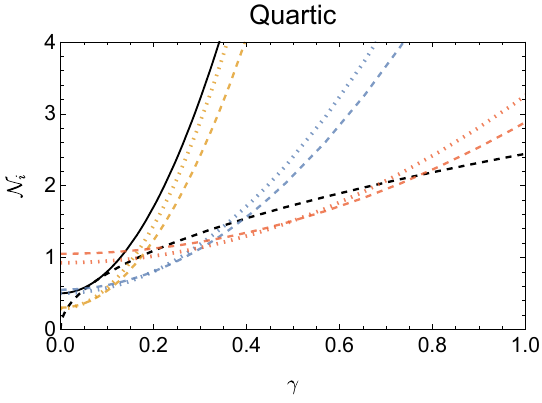}
    \caption{Nullifier variances at $\lambda=\gamma$ for loss $T=0.85$ (dashed curves) and thermalisation $\nb=0.05$ (dotted curves), for position squeezing (red), momentum squeezing (yellow), or no squeezing (blue). Nullifier variances below the ground state variance (black) and squeezed state threshold (dashed black) indicate nonclassicality. For the Gaussian sheared state (quadratic) momentum squeezing is always beneficial, and position squeezing always harmful, regardless of loss or thermalisation. The response to loss and thermalisation for the non-Gaussian cubic and quartic phase states depends on $\gamma$ in a similar fashion to that of the hypergraph states in Fig~\ref{LossATherm}. Additionally, for the cubic and quartic phase states the squeezed state limit is in fact a quantum non-Gaussianity criterion~\cite{moore_hierarchy_2022}.}
    \label{LossAThermSingleMode}
\end{figure}

For the nonlinear phase states $e^{i\gamma q^k}\ket{r}$ the equivalent of the hypergraph nullifiers are the nonlinear variables $p+k\lambda q^{k-1}$. For the states produced by the Gaussian shear gate $(k=2)$, loss and thermalisation can always be compensated by increasing the momentum squeezing, although we note that for this special case that the figure of merit (squeezing) overlaps with the quantity used to improve the figure of merit (initial squeezing). The same is not true for the nonlinear phase states, as has been shown for loss for the cubic phase state $(k=3)$~\cite{kala_cubic_2022}. Fig.~\ref{LossAThermSingleMode} shows that the optimisation of initial squeezing required for tolerance to loss and thermalisation for the hypergraph nullifiers also occurs for the nonlinear phase states. Therefore, the conclusions regarding preparing hypergraph nonclassicality in the presence of loss or thermalisation also apply to the nonclassicality of the nonlinear phase states. In fact, the stronger threshold for local squeezing used for hypergraph nonclassicality translates, in the single mode case, to an even stronger quantum non-Gaussianity criterion. 

\begin{figure}[h!]
        \includegraphics[width=0.24\columnwidth]{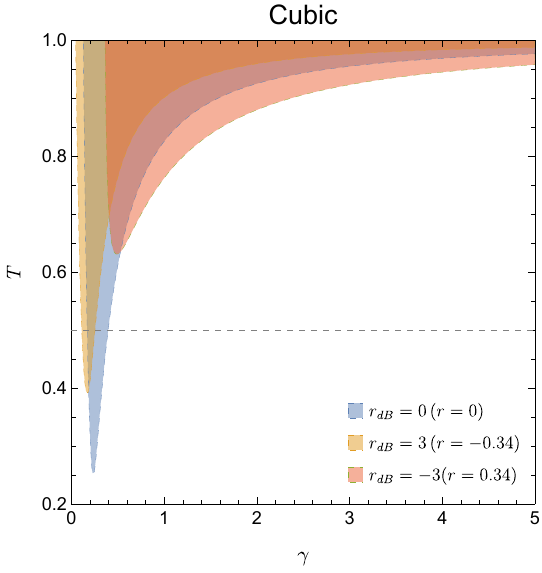}
        \includegraphics[width=0.24\columnwidth]{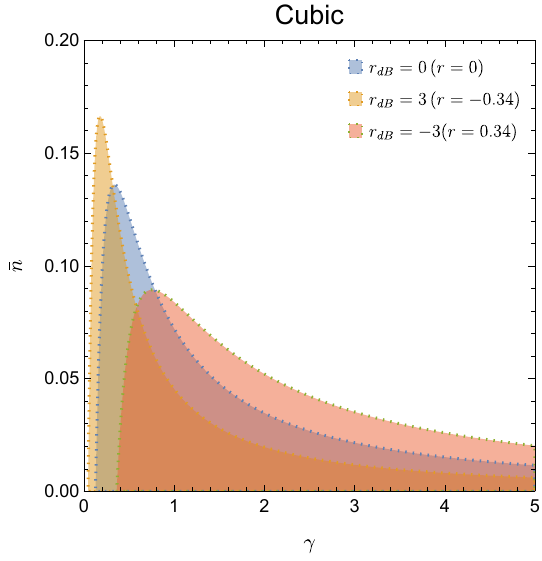}
        \includegraphics[width=0.24\columnwidth]{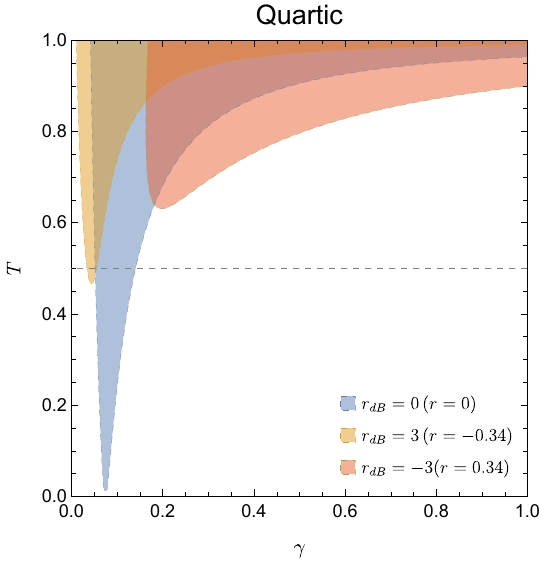}
        \includegraphics[width=0.24\columnwidth]{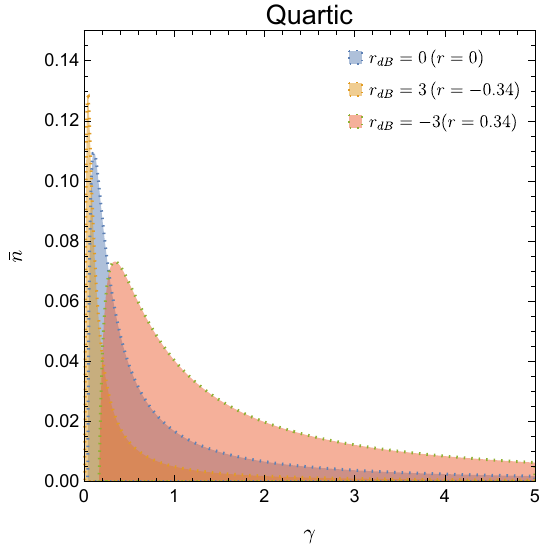}
    \caption{The maximum loss $T$ and thermalisation $\bar{n}$ against the nonlinearity $\gamma$ for various values of initial squeezing $r$. The states are the fully degenerate hypergraph triad and tetrad, which are the cubic and quartic phase gates respectively. Increasing the nonlinearity beyond the optimal value increases the sensitivity to loss. For values of $\gamma$ lower than this optimal value, loss can be partially compensated for by the initial squeezing. However for larger values of $\gamma$, squeezing also increases the sensitivity to loss. These are exactly the qualitative properties of the hypergraph state's sensitivity to loss.}\label{LossSM}
\end{figure}

In addition we show that qualitatively similar effects for the maximal loss or thermalisation also occur for the single mode nonlinear phase states. Fig.~\ref{LossSM} shows how these vary with $\gamma$ while still showing nonclassicality below the level of any initial squeezed state. 

\begin{figure}
    \centering
    \includegraphics[width=0.35\columnwidth]{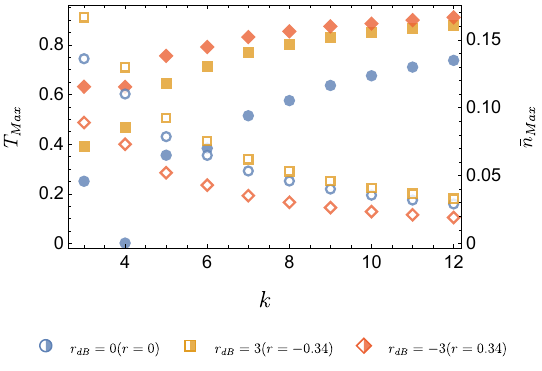}
    \caption{Maximum loss $T_{Max}$ (full plot markers) and  maximum thermalisation $\bar{n}_{Max}$ (empty plot markers) \textit{vs} single mode nonlinearity degree $k$, for momentum or position squeezing.}
    \label{SingleModeNonlinearKLoss}
\end{figure}

Finally we also show how these effects scale with the order of the nonlinearity for the nonlinear phase states in Fig.~\ref{SingleModeNonlinearKLoss}. The property that no squeezing is optimal for loss and position squeezing is optimal for thermalisation remains for single-mode nonlinearities. 

\end{appendix}


\end{document}